\documentclass[10pt]{article}
\usepackage{bbm}
\usepackage[final]{graphics}
\usepackage{amsmath}
\usepackage{amsfonts,amsbsy}
\usepackage{amssymb}

\catcode`\@=11

\newcount\@tempcntc
\def\@citex[#1]#2{\if@filesw\immediate\write\@auxout{\string\citation{#2}}\fi
  \@tempcnta\z@\@tempcntb\m@ne\def\@citea{}\@cite{%
        \@for\@citeb:=#2\do%
    {\@ifundefined{b@\@citeb}%
        {\@citeo\@tempcntb\m@ne\@citea%
                \def\@citea{,\penalty\@m\ }{\bf ?}\@warning%
                {Citation `\@citeb' on page \thepage \space undefined}}%
        {\setbox\z@\hbox{\global\@tempcntc0\csname b@\@citeb\endcsname\relax}
     \ifnum\@tempcntc=\z@ \@citeo\@tempcntb\m@ne%
       \@citea\def\@citea{,\penalty\@m}%
       \hbox{\csname b@\@citeb\endcsname}%
     \else%
      \advance\@tempcntb\@ne%
      \ifnum\@tempcntb=\@tempcntc%
      \else\advance\@tempcntb\m@ne\@citeo%
      \@tempcnta\@tempcntc\@tempcntb\@tempcntc\fi\fi}}\@citeo}{#1}}%

\def\@citeo{\ifnum\@tempcnta>\@tempcntb\else\@citea
  \def\@citea{,\penalty\@m}%
  \ifnum\@tempcnta=\@tempcntb\the\@tempcnta\else
   {\advance\@tempcnta\@ne\ifnum\@tempcnta=\@tempcntb \else
\def\@citea{--}\fi
    \advance\@tempcnta\m@ne\the\@tempcnta\@citea\the\@tempcntb}\fi\fi}

\catcode`\@=12

\begin{document}

\date{}
\title{
Long range correlations in high multiplicity hadron collisions:building bridges with ridges}
\author{Raju Venugopalan}
\maketitle
\begin{center}
Physics Department, 
  Brookhaven National Laboratory,
  Upton, NY, USA
\end{center}


\begin{abstract}
We discuss the physics of the ridge -- azimuthally collimated long range rapidity correlations -- in high multiplicity proton-proton and proton-nucleus collisions. We outline some of the theoretical discussions in the literature that address the systematics of these ridge correlations. 

\end{abstract}

\section{Introduction}

Quantum Chromodynamics (QCD) was established to be the right theory of the strong interaction shortly after the discovery of asymptotic freedom. In the 40 years since, the predictions of perturbative QCD have become so precise that they are used as background for new physics beyond the standard model. Nevertheless, if compared to Quantum Electrodynamics (QED) as a benchmark, there is a long way to go. In QED, simple questions about how electrons and photons interact with media are well understood; more generally, the collective properties of QED form a major part of what we call condensed matter physics. In QCD, we are a long ways away from this level of understanding. Even formulating the right experiments to address questions, for example, about how quarks and gluons scatter off an extended strongly interacting medium, is challenging. 

Prof. Wit Busza has been in the forefront of performing high energy scattering experiments that seek to address the elementary yet profound questions that bring into sharp relief our limitations in understanding nature's strong interaction. He realized early on that colliding protons at high energies off light and heavy nuclei are interesting in their own right~\cite{Busza:1975te}. They provide important insight into the multiple scattering of partons in a QCD medium, the energy loss and ``stopping" of fast partons in-media~\cite{Busza:1983rj}, and tackle the important question of how partons hadronize in and out of media. In several cases, the data pointed to intriguingly simple scaling patterns, which are still not completely understood~\cite{Busza}. 

With the advent of the Relativistic Heavy Ion Collider (RHIC), Wit Busza lead the PHOBOS experiment to efficiently produce first results from RHIC~\cite{Back:2000gw}. Proton-nucleus (or in this case deuteron-nucleus) collisions came to be seen primarily as a benchmark for patterns in nucleus-nucleus collisions. There were however results obtained in deuteron-gold collisions of strong shadowing effects in forward single inclusive hadron production~\cite{Arsene:2004ux} and in ``forward-forward" di-hadron correlations~\cite{dA-expt}. Both of these effects were {\it predicted} to arise as a  result of gluon saturation in nuclei~\cite{Kharzeev:2003wz,Marquet:2007vb}.

Proton-nucleus collisions at the LHC, with previously unimaginable center-of-mass energies of $\sqrt{s}=5020$ GeV/nucleon, have breathed fresh life into the subject. They have made transparent Wit Busza's view that these collisions are not merely a benchmark for interesting physics but are profoundly interesting in themselves in what they may offer of a deeper understanding of dynamics in QCD. In the rest of this talk, I will focus on the remarkable ridge phenomenon in high multiplicity proton-proton and proton-nucleus collisions. These  are two particle correlations that are long range in their relative psuedorapidity separation and collimated in relative azimuthal angle. They were first discovered at RHIC in A+A collisions~\cite{RHIC-ridge}; in particular, the PHOBOS experiment demonstrated correlations over 4 units in relative rapidity~\cite{PHOBOS-longrange}. At the LHC, the CMS heavy ion group, including several former PHOBOS members, have performed seminal work in uncovering such correlations in smaller sized systems. Recent reviews of the ridge effect can be found in \cite{Wang:2013qca,Loizides:2013nka,Li:2012hc,Kovner:2012jm}. 

\section{Initial state effects and the ridge}
\label{sec:initial}

In a series of three recent papers with Kevin Dusling~\cite{Dusling:2013oia,Dusling:2012wy,Dusling:2012cg}, we argued that  the ridge data  on two particle correlations in high multiplicity proton-proton and proton-nucleus collisions from the CMS collaboration~\cite{Khachatryan:2010gv,CMS:2012qk} provided strong evidence for gluon saturation and the Color Glass Condensate (CGC) effective field theory (EFT)~\cite{Gelis:2010nm}. 

Typically in perturbative QCD the only two parton correlation that one obtains is the back-to-back $\Delta\Phi\sim \pi$ correlation from the di-jet graph; this explains why none of the event generators saw the ridge collimation at $\Delta\Phi\sim 0$. In the CGC effective theory, the nearside $\Delta\Phi\sim 0$ collimation is obtained from connected two gluon production QCD graphs called ``Glasma graphs". They are QCD interference graphs. In conventional perturbative QCD computations, these graphs are ignored for good reason. Their contribution at high $p_T$ and in peripheral collisions is negligibly small. 

\begin{figure}[htbp]
\begin{center}
\resizebox*{6cm}{!}{{\includegraphics{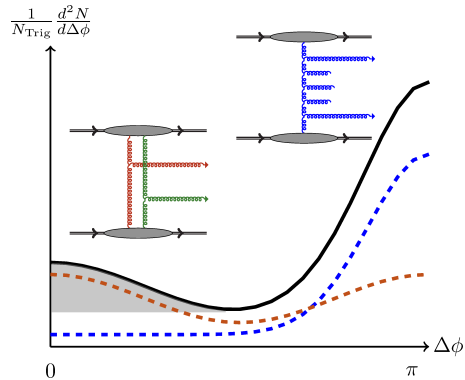}}}
\end{center}
\caption{\label{fig:graph} Sketch of initial state contributions to the di-hadron per trigger yield. The glasma graph on the left illustrates its contribution to the double inclusive cross-section (dashed orange curve). On the right is the 
back-to-back graph and the shape of its yield (dashed blue curve). The grey blobs denote emissions all the way from beam rapidities to those of the triggered gluons. The solid black curve represents the sum of contributions from glasma and back-to-back graphs. The shaded region represents the Associated Yield (AY) calculated using the zero-yield-at-minimum (ZYAM) procedure. Figure from ref.~\cite{Dusling:2012cg}. }
\end{figure}

However,  most remarkably, the high occupancy of gluons (for transverse momenta $k_\perp \leq Q_S$, where $Q_S$ is the saturation scale) in rare high multiplicity proton-proton events enhances such graphs by $\alpha_S^{-8}$. This corresponds to a strikingly large enhancement of $\sim 10^5$ for typical values of the probed QCD fine structure constant $\alpha_S$! Thus in the power counting of the EFT, gluon saturation ensures that Glasma graphs provide a significant  additional contribution in high multiplicity events to ``di-jet" QCD graphs. 

The importance of Glasma graphs was first discussed in ~\cite{Dumitru:2008wn} and the formalism developed in ~\cite{Gelis:2008sz,Dusling:2009ni}. It was first postulated as an explanation of the high multiplicity CMS proton-proton ridge in \cite{Dumitru:2010iy}, and a quantitative description of the nearside collimated yield obtained in \cite{Dusling:2012iga}. 

The di-jet contribution that is long range in rapidity is described in the CGC EFT by BFKL dynamics~\cite{Balitsky:1978ic,Kuraev:1977fs}. We showed in \cite{Dusling:2012cg} that BFKL dynamics does well in describing the awayside spectra in high multiplicity proton-proton collisions. The description is significantly better than PYTHIA-8~\cite{Khachatryan:2010gv}, and $2\rightarrow 4$ QCD graphs in the Quasi--Multi--Regge--Kinematics (QMRK)~\cite{Fadin:1996zv,Leonidov:1999nc}. Both of these approaches overestimate the awayside yield especially at larger momenta. 

Strictly speaking, the BFKL ``impact factors" are 
replaced by the gluon unintegrated distributions obtained from solutions of the running coupling Balitsky-Kovchegov equation~\cite{Balitsky:1995ub,Kovchegov:1999yj} . The latter encodes saturation effects in  small $x$ evolution. It is important to note that the corresponding di-jet contributions are then also enhanced $O(\alpha_S^{-4})$ by gluon saturation in high multiplicity hadron collisions. Without this saturation generated enhancement, the di-jet per trigger yield would not agree with the data. Subsequently we showed for the same parameter set as in proton-proton collisions, a good description is obtained for proton-nucleus collisions within the uncertainties of model parameters. The only difference between p+p and p+A collisions in our approach are the saturation scales (of projectile and target) chosen at the initial rapidity scale for small $x$ evolution. 

In fig.~(\ref{fig:graph}), we show a schematic sketch from refs.~\cite{Dusling:2013oia,Dusling:2012wy,Dusling:2012cg} of the Glasma and BFKL graphs. As shown in the figure, the Glasma graphs give a collimated contribution in $\Delta \phi$ that is mirror symmetric about $\Delta \phi = \pi/2$. The di-jet contribution is peaked back-to-back around $\Delta \phi = \pi$ and gives a negligible contribution on the nearside at $\Delta\phi=0$. 

\begin{figure}
\begin{center}
\resizebox*{6cm}{!}{{\includegraphics{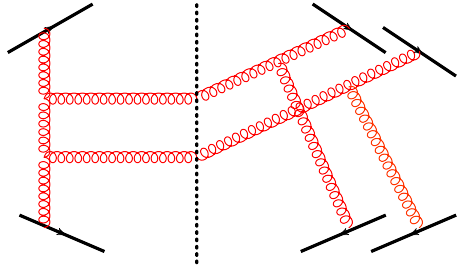}}}
\end{center}
\caption{\label{fig:interference} Sketch of typical quantum interference contribution between Glasma and BFKL graphs. The dashed line 
demarcates the separation between amplitude and complex conjugate amplitude. The solid lines represent color sources in the projectile and target. The sources in each are connected via a three point correlator to form a 
connected QCD diagram.}
\end{figure}

The Glasma+BFKL CGC framework describes reasonably well the associated yield per trigger obtained in p+Pb data at $\sqrt{s}=5.02$ TeV/nucleon by the CMS collaboration~\cite{CMS:2012qk}. Subsequently, both the ALICE~\cite{Abelev:2012ola} and
ATLAS~\cite{Aad:2012gla} collaborations published their di-hadron correlation results from the first LHC p+Pb run. The ALICE experiment
has an acceptance in $\Delta \eta$ of $|\Delta \eta|<1.8$, while the ATLAS experiment has an acceptance of $2 < |\Delta \eta|< 5$, close to the CMS acceptance of $2<|\Delta \eta|<4$.  In addition to the LHC results, the PHENIX collaboration at RHIC reanalyzed their deuteron-gold data at 200 GeV/nucleon and extracted a ridge signal in very central events~\cite{Adare:2013piz}.  All three of the experiments show that when the two particle yield in peripheral collisions is subtracted from the central events, a dipole structure remains that is long range in rapidity.  This is precisely what  we anticipated in our Glasma+BFKL graph scenario. Specifically, this is because the BFKL di-jet contribution has a weak dependence on centrality and the net Glasma graph contribution is symmetric around $\Delta\phi = \pi/2$. 

The LHC experiments have analyzed~\cite{Abelev:2012ola,Aad:2012gla,Chatrchyan:2013nka} the $v_n$ Fourier moments of the jet-subtracted di-hadron yields in proton-nucleus collisions\footnote{Note that this has not been done yet in proton-proton high multiplicity collisions because the jet yield is very large, potentially generating large systematic uncertainties in any such subtraction.}. They obtained a significant value of $v_2$ which approaches that observed in peripheral A+A collisions. Because of the $\Delta\Phi$ dependence of our Glasma graph contribution, we too would expect a significant $v_2$. However, the experiments observed a distinct $v_3$ contribution. In addition, $v_2\{4\}$, $v_2$ extracted from four particle cumulants, was significant. Both of these effects were widely interpreted as problematic for the initial state CGC picture. 

We consider this conclusion premature. In a systematic computation of the Glasma and di-jet graphs, there is also an interference contribution between the two. See fig.~\ref{fig:interference}. Previously, we had not included it because it is small. It is indeed so, but it could produce the needed $v_3$. Our preliminary conclusions are that this contribution gives a $v_3$ and that it has the right centrality and $p_T$ dependence seen in the data~\cite{Kevin}. However, interference contributions between the leading order Glasma graphs and the next-to-leading order di-jet graphs is subtle and further checks are necessary. Regarding $v_2\{4\}$, it has been conjectured that such an effect could possibly be obtained in an initial state scenario of overoccupied gluon fields~\cite{Dumitru:2013tja}. Detailed calculations though have yet to be performed.

\section{Final state effects and the ridge}

After ridges were seen in A+A collisions, there developed a consensus that the collimation of structures that were long range in rapidity occured due to radial hydrodynamic flow. More generally, it was observed that there was a strongly correlation between event geometry and $v_n$ moments~\cite{Alver:2010gr}.The observation of large ridges in proton-nucleus collisions that are comparable in size to those in peripheral nucleus-nucleus collisions have reinvigorated interpretations in terms of hydrodynamic flow. Occam's razor, sometimes called the principle of parsimony, is often invoked.  The reasoning goes as follows: If the p+A ridge looks like A+A, and the A+A is due to flow, then the result in p+A must also be due to flow. One might wish to consider though that the ``flow of the syllogism" could be in the reverse direction\footnote{Logicians understand well why the good friar wielded a razor and not an ax.}. After all, the ``looks like" part of the argument refers to peripheral A+A collisions. Additional force for the flow argument is adduced from the mass ordering of $v_2(p_T)$ for different hadron species~\cite{ABELEV:2013wsa}. In p+A collisions, it looks remarkably similar to that in peripheral A+A collisions, both in shape and magnitude. 

Computations in a hydrodynamical model reproduced a number of features of the proton-lead data~\cite{Bozek:2012gr,Bozek:2013df,Bozek:2013uha,Bozek:2013ska}. Other recent hydrodynamics based discussions of the ridge in p+A collisions can be found in refs.~\cite{Shuryak:2013ke,Qin:2013bha,Werner:2013ipa}. However, hydrodynamics when applied to p+A collisions is very sensitive to assumptions about the dynamics of the initial state and values of $\eta/s$.  The models cited involve such radically different assumptions about the multiparticle production in the initial state that one might almost believe they come from different underlying theories!

Our foray into a hydrodynamic description of small size systems was performed in the framework of the IP-Glasma+MUSIC model. The initial conditions in the IP-Glasma model are the IP-Sat CGC based initial conditions and include non-equilibrium 2+1-D Yang-Mills dynamics~\cite{Schenke:2012wb,Schenke:2012hg}. This evolution is subsequently matched event-by-event to viscous hydrodynamic evolution  described by MUSIC~\cite{Schenke:2010rr,Gale:2013da}.  The model gives excellent fits to data at both RHIC and LHC in A+A collisions~\cite{Gale:2012rq}. However, when applied to p+A collisions, the values of $v_2$ and $v_3$ are factors of 2-3 smaller than the data for the same values of $\eta/s$ that describe the A+A data~\cite{Bzdak:2013zma}. 

To help distinguish between final state models alone, it thus becomes essential to perform `apples-to-apples' comparisons. For instance, when it is stated that `p+A looks like A+A', what is meant is that $3\%$ central p+Pb collisions are being compared to very peripheral ($70$\% centrality)
Pb+Pb events. The most central events in CMS's comparison~\cite{Chatrchyan:2013nka} are $55$\% centrality in Pb+Pb--the same multiplicities correspond to 1 in 10 million events in p+Pb! 

Very recently, we performed a comprehensive study of single inclusive multiplicities and multiplicity distributions in p+p, p/d+A and A+A collisions within the IP-Glasma framework~\cite{Schenke:2013dpa}. With the results of this study, flow generated from the initial state configurations that generate the $70$\% central A+A collisions can be compared directly to the flow generated from initial state configurations corresponding to $3$\% central events in p+Pb collisions. Several models include the negative binomial distributions that reproduce multiplicity distributions (corresponding to centrality classes)  However thus far it is only in the IP-Glasma model that centrality classes are directly correlated with both the magnitudes and spatial distributions of gluon field configurations.

\section{Pandora's box}

So what's the physics that generates these ridge-like structures? Is it initial state dynamics, or collective flow, or some combination of the two? Is the physics due to long range gluon entanglement as suggested by the quantum inteference represented by the Glasma graphs? Or is it the smallest scale system in nature to exhibit collective dynamics? If so, how do we interpret such collective dynamics. Is the proton that collides in p+A collisions an unusually ``fat" proton~\cite{Coleman-Smith:2013rla} or a transversely aligned flux tube structure~\cite{Bjorken:2013boa} ? Even more exotically, is the dynamics reflecting an exploding sphaleron~\cite{Shuryak:2013sra} ? Another interesting idea is that the higher particle correlations may be due to a ``Glasmion" or Bose Condensation~\cite{Dumitru:2013tja}. The ideas outlined are not by any means fully  representative of the discussion in the literature.

On the plus side, the wide range of ideas suggest that the experimental results have attracted great interest. If any of these ideas prove to be correct, we will have learnt something very non-trivial about hadron structure and QCD dynamics from these high multiplicity proton-proton and proton/deuteron-nucleus collisions. 
Less charitably, this plethora of ideas may seem a little bewildering unless there were some way to tell them apart. We shall discuss here open issues that, from my perspective, each of these approaches has to resolve. 

\subsection{Open issues in the initial state+glasma scenario}

Let's first consider the Glasma+BFKL scenario. As outlined in section~\ref{sec:initial}, this approach is highly developed with quantitative comparisons to data in both proton-proton and proton-nucleus collisions. One problem this approach faced was with $v_3$. In section~\ref{sec:initial}, we noted that there is an interference contribution that a) has to be there, and b) was previously neglected. Clearly this needs to be fleshed out. Another open issue is whether one can get a $v_2\{4\}$ that is significant and has the systematics seen in data. Finally for this approach to be successful, one has to understand why one sees in proton-nucleus collisions the same mass ordering patterns~\cite{ABELEV:2013wsa} and even the same magnitudes seen in peripheral A+A collisions. While the intial state dynamics does not know about hadronization, it appears that the hadronization patterns seen in the data must be universal vacuum properties for initial state physics to be the primary cause of the ridge. 

The theoretical status of the initial state models is not fully satisfactory and an improved treatment of these could hold the answer to apparent puzzles. The computations of the Glasma graphs in \cite{Dusling:2013oia,Dusling:2012wy,Dusling:2012cg} are carried out in the ``dense-dense" limit, where both projectile and target color sources are ``saturated"--the gluon fields have maximal occupancy. In my view, for the rare events studied, the dense-dense framework is the appropriate one. The dynamics in this framework includes both coherent multiple scattering and small $x$ QCD evolution from both the projectile and the target. However the computations in \cite{Dusling:2013oia,Dusling:2012wy,Dusling:2012cg} are performed within a $k_T$ factorization approximation where coherent multiple scattering effects are not fully included. For $p_T\sim Q_S$, where the ridge yield is largest, these effects may be relatively small. Nevertheless, this systematic uncertainty has to be quantified. Computations have also been performed in the other limit--significant coherent multiple scattering but no small $x$ QCD evolution. These computations were performed 
numerically in \cite{Lappi:2009xa}--and extended signficantly recently in the IP-Glasma model~\cite{Bjoern}. In principle, as long as $\Delta\eta < 1/\alpha_S$, one knows how to include both coherent multiple scattering and QCD evolution effects on di-gluon production in dense-dense framework~\cite{Gelis:2008sz}. This matter is under active investigation\footnote{In the dilute-dense limit, where the projectile 
proton is at large $x$ and the nuclear target is dense, analytical computations of multiple scattering  and QCD evolution contributions  are feasible for $\Delta\eta <1/\alpha_S$~\cite{Kovchegov:2012nd,Kovchegov:2013ewa}.  A numerical algorithm has been  proposed~\cite{Iancu:2013uva}  for the more challenging $\Delta \eta >1/\alpha_S$ case. Unfortunately the dilute-dense framework is not appropriate for the ridge kinematics at the LHC.}. The di-jet contribution is also approximated in the treatment of \cite{Dusling:2013oia,Dusling:2012wy,Dusling:2012cg}. An improved treatment is extremely challenging and beyond the scope of present studies. 

As noted, the full CGC framework also includes coherent multiple scattering. This complicates a naive initial state/final state dichotomy. Such effects correspond to the non-equilibrium dynamics of coherent classical fields in the Glasma, which are incorporated in the IP-Glasma model. In that context, whether one has ``final" state effects would correspond quantitatively to the effect of ``matching" the Glasma to hydrodynamics. The IP-Glasma performs this matching imperfectly and there is likely significantly more rescattering in the non-equilibrium regime than considered previously. In this respect, there has been significant recent progress in performing 3+1-D Yang-Mills simulations that encode the non-trivial Glasma dynamics~\cite{Berges:2013eia,Epelbaum:2013waa,Gelis:2013rba,Berges:2013fga}. The ideas advocated in 
\cite{Dumitru:2013tja} have to be viewed in this context. We note that recently the ALICE collaboration extracted a coherent fraction of $22\pm 12$\% for charged soft pion emission in Pb+Pb collisions~\cite{Abelev:2013pqa}. This emission is observed to be very weakly dependent on centrality.

\subsection{Panta Rhei?}

The final state\footnote{Despite the terminology, there is no such thing as a purely final state model. All hydrodynamic models include as input assumptions regarding the dynamics of two and higher point correlations.} models \cite{Bozek:2012gr,Bozek:2013df,Bozek:2013uha,Bozek:2013ska} of the ridge do not include non-equilibrium dynamics. Even in a ${\cal N}=4$ supersymmetric non-Abelian plasma at strong coupling, gauge-gravity duality inspired computations suggest that viscous hydrodynamics is not applicable for the first half fermi or so; even when  viscous hydrodynamics becomes applicable, viscous effects are considerable at early times in heavy ion collisions~\cite{Chesler:2010bi}. For small size systems, these should matter even more. 

The large quantitative differences seen between different hydrodynamic initial state scenarios for identical values of $\eta/s$ suggest that the final state problem is an initial state problem after all. One gets out what one puts in. This is qualitatively different from central heavy ion collisions where different initial conditions, for fixed $\eta/s$, give results that vary by $\sim 15$\%. The focus for small sized systems should therefore be on models of the initial state. 

In this regard, let us briefly consider the aforementioned ``non-Glasma" models. The model of Shuryak and Zahed~\cite{Shuryak:2013sra} is closer to the Glasma picture than one might at first think. Their formuation is within Reggeon Field Theory; there is a large body of work that demonstrates a close correspondence between this framework and the CGC--see for instance ~\cite{Altinoluk:2013rua,Caron-Huot:2013fea} and references therein. In particular, the Mueller-Kancheli diagrams considered correspond to the Glasma and di-jet diagrams in the CGC. The (important) differences though are in how these are computed. In the CGC, the semi-hard scale $Q_S$ allows for weak coupling computations; the Shuryak-Zahed scenario is completely non-perturbative. It therefore relies on the gauge-gravity duality to perform computations. A qualitative difference to the weak coupling Glasma framework is that the former has ``critical" onset of the ridge like multiparticle correlations as the multiplicity of events increases. One consequence is that radial flow is very strong and stronger even in proton-proton collisions than in proton-nucleus collisions. An open question for this picture is why then 
the ridge is ~4 times larger in p+A collisions than in p+p collisions. Further, in this non-perturbative scenario, one should expect very significant quenching of mini-jets in p+p collisions~\cite{Zakharov:2013gya}. None is seen. On the contrary, short distance physics is alive and kicking hard--the jet contribution is huge. 

A conjecture as to why the ridge may be large in p+A collisions is the ``fat proton" proposal in \cite{Coleman-Smith:2013rla}. In this scenario, the proton has to be 3 fermi in size. To understand this conceptually, imagine a proton at low energies consisting of valence partons, a gluon or two, and pions. As the proton is boosted to high energies, higher Fock states with large number of gluons appear. However, because of confinement, these have 
to be localized primarily in the center of the nucleon; Gribov diffusion suggests that the proton radius squared can grow at most as a logarithm of the energy. Producing pions gets around color confinement. However as is well known from the lore of multiperipheral ladders, these contributions are suppressed with energy. Conceptual issues aside, the proton-nucleus collisions where the ridge is significant are not especially rare. Further, if such large spatial Fock configurations are seen in proton-nucleus collisions, one should see even larger configurations in the extremely rare proton-proton collisions that generate the much tinier proton-proton ridge. 

Atypical spatial Fock configurations are also considered in \cite{Bjorken:2013boa}. The focus here is proton-proton collisions, where the ridge configurations are truly rare. The configurations conjectured are elongated in 
the transverse plane (though with sizes $\sim 1$ fermi) possessing large eccentricity, which can translate into large elliptic flow\footnote{It is a little unclear if flow is an intended consequence of their scenario.}.  This scenario too does not explain what happens to mini-jets in the presence of strong flow. Further pertinent questions include whether $v_3$ is generated, and the relevance of the scenario for the ridge seen in proton-nucleus  collisions. 

We conclude with a brief discussion of results that are not easily reconciled in any scenario. As discussed, the mass ordering of $v_2$ seen in proton-nucleus collisions is seen by some as strong evidence of collectivity~\cite{Bozek:2013ska}. However mass ordering in $v_2$ is strongly tied to radial flow. Mass splitting is seen in $\langle p_T\rangle$ for events with $N_{\rm ch.} =5$~\cite{Abelev:2013haa}. Its a stretch to interpret such events to be ``collective" and PYTHIA with color reconnections reproduces the data well. Many aspects of the striking patterns in hadrochemistry are puzzling. 
Another puzzle concerns the identical values of $v_3$ in Pb+Pb collisions as a function of $N_{\rm charge}$ (corresponding to variations in the $55$\%-$92$\% centrality window)  when compared to $v_3$ in the same $N_{\rm charge}$ range in p+Pb collisions (corresponding to variations ranging from $3$\% central to one in $10^{7}$ events). Likewise, the mass ordering and magnitude of $v_2$ is again nearly identical in p+A and peripheral A+A collisions. There is {\it a priori} no reason to believe that systems with such vastly different properties should display such uniform patterns.

\section{Summary}

Wit Busza could perhaps not have foreseen 40 years ago that his pioneering studies would lead to such creative ferment in experiment and theory. As he has emphasized time and again, the data show patterns across energies, particle species, and system size, that appear much simpler and robust than predicted by theoretical models. This is indeed true. Nevertheless, I believe we have learned a great deal. The ability of experiment to falsify/help refine theoretical models is much greater than ever. Theoretical models are more sophisticated and more faithful to the underlying theory than previously. I am hopeful that with Wit's continued creative input lasting progress can be achieved.

\section*{Acknowledgements}
R.V's research was supported by DOE Contract No.
DE-AC02-98CH10886. He is grateful to James Bjorken, Adam Bzdak, Adrian Dumitru, Kevin Dusling, Dhevan Gangadharan, Jan Fiete Grosse-Oetringhaus, Miklos Gyulassy, Martin Hentschinski, Edmond Iancu, Yuri Kovchegov, Wei Li, Constantin Loizides, Larry McLerran, Peter Petreczky, Bjoern Schenke, Anne Sickles, Derek Teaney, Giorgio Torrieri, Prithwish Tribedy and Konrad Tymoniuk for useful discussions on the topics discussed here. He thanks W. Busza, A. Bzdak, K. Dusling, L. McLerran and B. Schenke for a close reading of the manuscript.


\end{document}